\documentclass[aps,twocolumn, amsmath,amssymb,pra,superscriptaddress]{revtex4-2}

\usepackage{graphicx}
\usepackage[dvipsnames]{xcolor}
\usepackage{bm}

\usepackage{empheq,etoolbox}
\usepackage[makeroom]{cancel}
\usepackage{bbold}
\usepackage{comment}
\usepackage{appendix}

\usepackage{mathabx}
\usepackage{amsmath}
\usepackage{booktabs}

\usepackage[section]{placeins}

\usepackage{verbatim}
\usepackage{xcolor}
\definecolor{lightgrey}{rgb}{0.8, 0.8, 0.8}
\definecolor{misprint}{rgb}{0, 0, 1}
\definecolor{remark}{rgb}{0.0, 0.5, 0.69}

\newcommand{\changes}[1]{\textcolor{black}{#1}}

\definecolor{purple}{rgb}{0.8,0,0.6}
\definecolor{darkgreen}{rgb}{0.00,0.6,0.00}

\begin{document}

\title{Velocity dispersion profiles of dwarf spheroidal galaxies with self-interacting ultralight dark matter}
\author{K. Korshynska}
\affiliation{Institut für Mathematische Physik, Technische Universität Braunschweig, Mendelssohnstraße 3, 38106 Braunschweig, Germany}
\affiliation{Fundamentale Physik für Metrologie FPM, Physikalisch-Technische Bundesanstalt PTB, Bundesallee 100, 38116 Braunschweig, Germany}
\affiliation{Department of Physics, Taras Shevchenko National University of Kyiv, 
64/13, Volodymyrska Street, Kyiv 01601, Ukraine}
\author{E. V. Gorbar}
\affiliation{Department of Physics, Taras Shevchenko National University of Kyiv, 
64/13, Volodymyrska Street, Kyiv 01601, Ukraine}
\affiliation{Bogolyubov Institute for Theoretical Physics, 14-b Metrolohichna Street, Kyiv 03143, Ukraine}
\author{Y. M. Bidasyuk}
\affiliation{Bogolyubov Institute for Theoretical Physics, 14-b Metrolohichna Street, Kyiv 03143, Ukraine}
\author{A. I. Yakimenko}
\affiliation{Department of Physics, Taras Shevchenko National University of Kyiv,
64/13, Volodymyrska Street, Kyiv 01601, Ukraine}
\affiliation{Dipartimento di Fisica e Astronomia ’Galileo Galilei’,
Universit{\'a} di Padova, via Marzolo 8, 35131 Padova, Italy}
\author{Y. Revaz}
\affiliation{Laboratoire d’Astrophysique, EPFL, Observatoire de Sauverny, 1290 Versoix, Switzerland}



\begin{abstract}
Dark-matter-dominated dwarf galaxies provide an excellent laboratory for testing dark matter models at small scale and, in particular, the ultralight dark matter (ULDM) class of models. Within the framework of self-interacting bosonic dark matter, we use the observed velocity-dispersion profiles of seven dwarf spheroidal galaxies to constrain the parameters of ULDM. In our modeling, we account for the impact of the baryonic component on the velocity dispersion and ULDM halo structure. We find that the repulsive self-interaction of ULDM, which fits the observations, is almost negligible, consistent with non-interacting ULDM with a boson mass of approximately $1.6 \times 10^{-22}\,\mathrm{eV}$. In contrast, for attractively interacting ULDM, the best fit corresponds to a smaller boson mass of about $1.3 \times 10^{-22}\,\mathrm{eV}$, with self-interaction playing a significant role in shaping the dark-matter halo and thereby influencing the interpretation of observations. 
\end{abstract}

\maketitle

\section{Introduction}

Nature of dark matter (DM) has been an intriguing mystery for astrophysics, cosmology, and particle physics since nearly a century now \cite{Bertone_2018}. Its existence has been predicted from astrophysical observations and modeled using different theoretical assumptions about its nature. Usually, it is assumed that DM is a non-relativistic collisionless fluid, which gives rise to a cold dark matter (CDM) model \cite{blumenthal1984formation}. While this model explains many of the observed astrophysical phenomena, it encounters a few challenges, such as cusp-core tension \cite{weinberg2015cold, de_Blok_2009}. One way to resolve them is to improve the modeling of galaxies by accounting for the baryonic component, the presence of black holes, and other relevant physical processes \cite{sales2022baryonic}. Another possible solution is to assume that the dark matter has some small-scale structure being composed of ultralight bosons in the mass range $m \in [10^{-18}, 10^{-24}]$~eV and constituting the ultralight dark matter (ULDM) class of models \cite{Ferreira,jackson2023search, schive2025fuzzy}.  

\changes{The properties of an ULDM halo are very sensitive to the boson mass.} \changes{For instance, the ULDM with $m \gtrsim 10^{-20}$~eV behaves similarly to standard CDM on galactic scales with pronounced wave-like behavior only in a small inner region of the halo \cite{Ferreira}. This regime has been previously considered for the non-interacting ULDM (sometimes referred to as fuzzy DM)  \cite{goldstein2022viability,teodori2025ultralightdarkmattersimulations, dalal2022excluding, may2025updated}, aiming to constrain the boson mass from the observations of stellar dynamics.} \changes{In particular, using} the velocity dispersion measurements in Fornax, Sculptor, Draco, Sextans, Ursa Minor, and Carina dwarf galaxies, it was concluded that the mass of fuzzy bosonic DM particle should be larger than $10^{-20}$~eV \cite{goldstein2022viability}. \changes{While for the boson mass $\sim 10^{-20}$~eV the DM halo has the wave-like coherent structure within the coherent length smaller than $\sim$~kpc,
for a boson of smaller mass $\sim 10^{-22}$~eV the coherent length can be of the order of kpc. Such ULDM  introduces wave-like behavior on the galactic scales and, therefore, addresses small-scale challenges of CDM \cite{Ferreira, chavanis2019predictive}.} In this study, we use the velocity dispersion profile of a set of dwarf galaxies to put constraints on \changes{the wave-like ultralight} dark matter and its properties. Since the velocity dispersion curves have been proven to be an efficient tool to constrain non-interacting ULDM, in the present study we investigate the more general case of self-interacting ULDM \cite{lee1996galactic, chavanis2019predictive, Ferreira}. In this case, the local self-interaction of ultralight bosons may significantly modify the properties of an ULDM halo  \cite{chavanis2019massradius}.

\changes{For the boson mass $m\sim 10^{-22}$~eV, we can} consider ULDM to be in the state of Bose-Einstein condensate (BEC), implying that the density distribution of DM halo is given by the solution of the Gross-Pitaevskii-Poisson system of equations. The latter equations define the shape and radius $R$ of ULDM halo of given mass $M_\textrm{DM}$ and ULDM model parameters
\cite{chavanis2019massradius}. We focus on the ground state solution given by a solitonic core, which is believed to be a good model for ultracompact halos \cite{chavanis2019predictive}. Thus, for given ULDM parameters, we determine
the halo density profile, which defines the gravitational potential and the velocity dispersion, assuming the system to be at Jeans equilibrium. We fit this theoretical prediction to the observable line-of-sight velocity dispersion for seven Milky Way dwarf spheroidal (dSph) galaxies~\cite{supp, walker2007velocity}. These galaxies are particularly convenient targets to investigate the properties of DM since they are known to be DM-dominated \cite{pascale2025leo}.
They are characterized by the close-to-flat velocity dispersion curves and are supposed to have total gravitational masses of the order of $10^{8-9}M_\odot$ \cite{walker2007velocity}.
In our analysis, we use these observations to fit not only the masses of the dwarf DM haloes, but also to put constraints on the particle mass and the strength of self-interactions in ULDM.

The paper is organized as follows. In Sec.~II we recapitulate the density and gravitational potential of the baryonic component of the dSphs and introduce ULDM for the modeling of DM component. In Sec.~III we describe the theoretical model for the velocity dispersion and the fitting procedure. Our first results are shown in Sec.~IV, where we investigate the fitting of theory to observations while neglecting the gravitational potential of a small baryonic matter component. In Sec.~V we self-consistently account for the baryonic gravitational potential, which contributes to the velocity dispersion and modifies the shape of the ULDM halo. Our results for the fitted ULDM density profiles in the seven dSphs are shown and discussed in Sec.~VI, where we also analyze the role of the baryonic matter and ULDM self-interactions in our analysis. Our conclusions are made in Sec.~VII.

\section{Matter content of a dwarf galaxy}

In this section, we briefly recapitulate the main features of ULDM models and of baryonic matter in a spherically symmetric galaxy. We focus on the sample of seven dSphs \cite{walker2007velocity}.

\subsection{Baryonic matter content of dwarf galaxies}
\label{Subsec: Baryonic matter content of dwarf galaxies}

A simple model for the baryonic density and gravitational potential in dwarf galaxies is the Plummer sphere \cite{plummer1911problem}. It determines the density $\rho_\textrm{b}(r)$ and the gravitational potential $\Phi_\textrm{b} (r)$ of baryons as
\begin{eqnarray}
    \rho_\textrm{b}(r) &=& \frac{3 b^2 M_\textrm{b}}{4 \pi (r^2 + b^2)^{5/2}} \, ,\label{eq: rhob}\\
    \Phi_\textrm{b}(r) &=& - \frac{G M_\textrm{b}}{\sqrt{r^2 + b^2}} \, ,\label{eq: phib}
\end{eqnarray}
and is characterized by the two parameters - mass $M_\textrm{b}$ and Plummer radius $b$, where $1.3b$ is the half-mass radius \cite{hodson2020distribution}. 

\begingroup
\setlength{\tabcolsep}{10pt} 
\renewcommand{\arraystretch}{1.2} 
\begin{table}[h]
\begin{tabular}{ |c|c|c| } 
 \hline 
 Galaxy   & $M_\textrm{b}/M_{\odot}$ & $b$/pc\\  
 \hline
 Carina  &  $1.07 \times 10^6$ & 241\\ 
 Draco  & $2.9 \times 10^5$ & 196 \\ 
Fornax  & $3 \times 10^7$ & 668\\
 Leo I & $5.5 \times 10^6$ & 246\\
 Leo II  & $1.7 \times 10^6$ & 151\\
 Sculptor  & $3 \times 10^7$ & 260\\
 Sextans   & $4.1 \times 10^5$ & 682\\
 \hline
\end{tabular}
\caption{Parameters of baryonic matter, the total stellar mass $M_\textrm{b}$ and Plummer radius $b$ entering the Plummer distribution (\ref{eq: rhob}), in dwarf spheroidals inferred from observations (see Ref.~\cite{hodson2020distribution} for overview) as given in Refs.~\cite{cappellari2006sauron, kowalczyk2019schwarzschild} for Fornax, Ref.~\cite{de2014episodic} for Carina, Ref.~\cite{refId0,read2018case} for Draco, Ref.~\cite{pascale2025leoiclassicaldwarf} for Leo~I, Ref.~\cite{Koch_2007} for Leo~II, Ref.~\cite{battaglia2008kinematic} for Sculptor and Ref.~\cite{10.1111/j.1365-2966.2010.17745.x} for Sextans.}
\label{Table: baryonic parameters}
\end{table}
\endgroup

For the considered seven dwarf spheroidals, orbiting Milky Way, the parameters of baryonic matter are presented in Table~\ref{Table: baryonic parameters}. These parameters come from the observed stellar luminosity profiles and we use these data from Table~\ref{Table: baryonic parameters} as a model-independent observational input.

\subsection{Ultralight dark matter}
\label{Subsec: Ultralight dark matter}

By assuming ULDM class of models, we consider the DM density profiles, which correspond to the wavefunction $\psi$ of the ULDM. The wavefunction $\psi$ obeys the Gross-Pitaevskii-Poisson (GPP) equations
\begin{align} 
&  i\hbar \frac{\partial\psi}{\partial t} = \left[-\frac{\hbar^2}{2m}\nabla^{2} + g N|\psi|^{2} + m \left(\Phi_\textrm{DM} + \Phi_\textrm{b} \right)\right]\psi,
  \label{GP equation}\\
& \nabla^{2}\Phi_\textrm{DM} = 4\pi G M_\textrm{DM}|\psi|^{2}\, ,
  \label{Poisson equation}
\end{align}
where $m$ stands for the boson mass, $N$ is the number of bosons and $M_\textrm{DM} = Nm$ is the total ULDM mass. Here $g = 4\pi a_\textrm{s} \hbar^2/m$ is the coupling strength of the local self-interaction and $a_\textrm{s}$ is the s-wave scattering length which together with $m$ fully determines the considered ULDM model. The long-range interactions are defined by the gravitational potential of baryons (\ref{eq: phib}) and the self-induced ULDM gravitational potential $\Phi_\textrm{DM}$. The latter is related to the ULDM density $\rho_\textrm{DM} = M_\textrm{DM}|\psi|^{2}$ via the Poisson equation (\ref{Poisson equation}). 

In the paper, we assume the virialized equilibrium ULDM configuration, which corresponds to the ground state solution of GPP equations and can be well-approximated with the Gaussian ansatz \cite{chavanis2014self, chavanis2011mass}. The corresponding Gaussian density profile $\rho_\textrm{DM}(r)$ and the gravitational potential $\Phi_\textrm{DM}(r)$ read
\begin{eqnarray}
    \rho_\textrm{DM} (r) &=& M_\textrm{DM}\left(\frac{1}{\pi R^2}\right)^{3/2}e^{-r^2/R^2} \, \label{eq: DM density} ,\\
    \Phi_\textrm{DM}(r) &=& -G M_\textrm{DM} \frac{\mathrm{erf}(r/R)}{r} \label{eq: DM potential}\, ,
\end{eqnarray}
where $R$ defines the length scale of ULDM halo. The size of the halo, where $99\%$ of the ULDM mass is contained, can then be found as $R_{99} = 2.38R$ \cite{chavanis2014self}. Equations~(\ref{eq: DM density}) and (\ref{eq: DM potential}) provide a good approximation to the exact solitonic solution of GPP equations if the condition $G\changes{M_\textrm{DM}^2} m |a_\textrm{s}|/\hbar^2 \lesssim 1$ holds \cite{chavanis2014self}. This is always the case for attractively interacting ULDM $a_\textrm{s} < 0$ and is fulfilled for $a_\textrm{s} > 0$ if moderate interaction strengths are considered \cite{chavanis2014self}.

As shown in Ref.~\cite{chavanis2014self} the total energy $E = \Theta_Q + U + W$ corresponding to this solution consists of the three components: quantum kinetic energy $\Theta_{Q}$, internal energy $U$ (due to self-interaction), and gravitational energy $W$, which read
\begin{equation}
   \Theta_Q = \frac{3}{4} \frac{\hbar^2 M_\textrm{DM}}{m^2 R^2} ,\, U = \frac{a_\textrm{s} \hbar^2 M_\textrm{DM}^2}{\sqrt{2 \pi} m^3 R^3}, \, W = - \frac{GM_\textrm{DM}^2}{\sqrt{2\pi} R}\, . \label{eq: energy compnents}
\end{equation}
Imposing the condition that the energy $E$ of the ground state ULDM soliton has to be minimized, one can find a corresponding mass-radius relation \cite{chavanis2011mass}. In the absence of self-interaction $a_\textrm{s} = 0$, the hydrostatic equilibrium is achieved for $R = 3\sqrt{\pi}R_Q/\sqrt{2}$, where $R_Q = \hbar^2/(G M_\textrm{DM}m^2)$ \cite{chavanis2014self}, while accounting for $a_\textrm{s} \neq 0$ one obtains $R = R_\textrm{DM}$
\begin{equation}
    R_\textrm{DM} = \frac{3\sqrt{2\pi}}{4} R_Q \left(1 + \sqrt{1 + \frac{8}{3\pi} \frac{M_\textrm{DM}}{m}\frac{a_\textrm{s}}{R_Q}}\right).
    \label{eq: first mass-radius relation}
\end{equation}
In Ref.~\cite{chavanis2011mass} an additional negative branch of the mass-radius relation for $a_\textrm{s} < 0$ is also discussed. However, it corresponds to the unstable maximum of the soliton energy $E$ and, therefore, will not be considered in what follows.

 \begin{figure}[htp]
    \centering  \includegraphics[width=.45\textwidth]{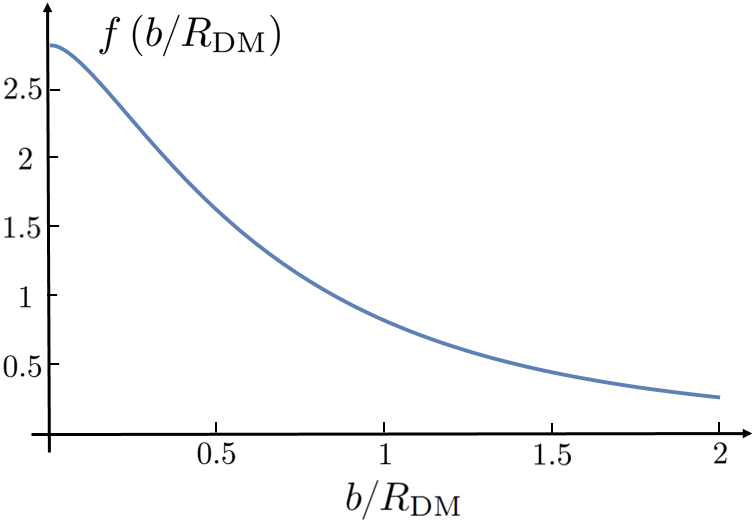}
    \caption{The function
    of $b/R_\textrm{DM}$ which defines the relative contribution of baryonic matter in Eq.~(\ref{eq: R(M) perturbed by bulge}).}
    \label{Fig: func of b/R}
\end{figure}

Note that if one accounts for the gravitational potential $\Phi_\textrm{b}$ induced by baryonic matter and given by Eq.~(\ref{eq: phib}), the total energy of ULDM acquires an additional component
\begin{eqnarray}
    W_\textrm{b} &=& 4 \pi \int_0^{+\infty} dr r^2 \rho_\textrm{DM}(r) \Phi_\textrm{b}(r) \\
    &=& - \frac{4G M_\textrm{DM} M_\textrm{b}}{\sqrt{\pi} R} \int_{0}^{+\infty}dx \frac{x^2 e^{-x^2}}{\sqrt{x^2 + (b/R)^2}} \, .
\end{eqnarray}
Due to this energy term the BEC deforms and, therefore, the mass-radius relation (\ref{eq: first mass-radius relation}) changes to
\begin{eqnarray}
    R &=& \frac{3 \sqrt{2\pi}}{4} \tilde{R}_Q \left(1 + \sqrt{1 + \frac{8}{3 \pi} \frac{M_\textrm{DM}}{m} \frac{a_\textrm{s}}{\tilde{R}_Q} }\right) \, , \label{eq: R(M) perturbed by bulge}\\
    \tilde{R}_Q &=& \frac{R_Q}{1 + M_\textrm{b}/M_\textrm{DM} \times f (b/R_\textrm{DM})} 
    \label{eq: R(M) perturbed by bulge 2}\, ,
\end{eqnarray}
where $\tilde{R}_Q$ is given by $R_Q$ modified due to the presence of baryonic matter. We see that due to additional gravitational attraction of baryonic matter, the size $R$ of ULDM decreases compared to the unperturbed case $R_\textrm{DM}$ (\ref{eq: first mass-radius relation}). The modification is defined by the ratio $M_\textrm{b}/M_\textrm{DM}$ between the masses of baryonic and DM components and also depends on the ratio of their sizes $b/R_\textrm{DM}$ via the function $f (b/R_\textrm{DM})$
\begin{eqnarray}     
f\left(\frac{b}{R_\textrm{DM}}\right) &=& \frac{3\sqrt{2\pi}}{2}\textrm{U}\left[3/2, 0, \left(\frac{b}{R_\textrm{DM}}\right)^2\right],\nonumber \\ \,
   \textrm{U}(\alpha, \beta, z) &=& \frac{1}{\Gamma(\alpha)}\int_0^{+\infty}dt \textrm{e}^{-zt} t^{\alpha - 1} (1+t)^{\beta-\alpha-1},
   \quad \label{eq: confluent U} 
\end{eqnarray}
where $U(\alpha, \beta, z)$ is the confluent hypergeometric function and $f (b/R_\textrm{DM})$ is illustrated in Fig.~\ref{Fig: func of b/R}. The derivation of results (\ref{eq: R(M) perturbed by bulge}-\ref{eq: confluent U}) is given in Appendix~\ref{Appendix: Mass-radius relation for ULDM with baryons}.

Note that we model an ULDM halo as a coherent BEC soliton (\ref{eq: DM density}) and neglect the possible presence of the density tail, which is expected to appear on the outskirts of the virialized solution of the GPP equations \cite{schive2014understanding}. It forms an isothermal envelope around the BEC core and can be described by the Navarro-Frenk-White (NFW) density profile \cite{schive2014understanding, chavanis2019predictive}. This isothermal envelope is believed to be negligibly small for the considered dSphs because they are ultracompact according to the classification of Ref.~\cite{chavanis2019predictive}, i.e., have masses $\sim 10^8 M_\odot$ and radii $\sim 1$~kpc.  Moreover, the isothermal envelopes of dwarfs, if present, would be hardly distinguishable from the isothermal envelope of the host Milky Way halo. The latter is believed to have most of its DM mass in the envelope and a rather small $\sim 1$~kpc core in the very center \cite{chavanis2019predictive}. Further analysis in this direction would require a complex modeling of the system of the host halo and the satellite haloes, cf. Refs.~\cite{gorkavenko2024dynamical, gorkavenko2024dynamicalinrot}, and goes beyond the scope of the present study.

\section{Velocity dispersion and fitting procedure}

\subsection{Velocity dispersion}

Let us consider the observed dynamics of baryonic matter with density $\rho_\textrm{b}(r)$ in the gravitational field $\Phi(r)$ of the matter content of a galaxy: ULDM and baryonic matter. In a galaxy, stars are described by a probability to be found in the infinitesimal space $[x,x+dx]$ and velocity $[v,v+dv]$ intervals. Assuming the system to be collisionless, this probability obeys the collisionless Boltzmann equation \cite{binney2011galactic}. If the system is at equilibrium, one can derive the Jeans equations for the radial velocity dispersion
\begin{equation}
    \overline{v_r^2} = \frac{1}{\rho_\textrm{b}(r)} \int_r^\infty dr' \rho_\textrm{b}(r') \frac{d\Phi}{dr'}\, ,
    \label{eq: dispersion general formula}
\end{equation}
where the ergodic distribution function was assumed, implying that the velocity dispersion is isotropic, i.e., $\overline{v_\theta^2} = \overline{v_\phi^2} = \overline{v_r^2}$. The observed velocity dispersion of the seven dSphs can be either isotropic or have a small anisotropy (see cored fit of the dSphs in Ref.~\cite{walker2009universal}), in what follows we will focus on the isotropic case given by Eq.~(\ref{eq: dispersion general formula}).

In Eq.~(\ref{eq: dispersion general formula}) the baryonic matter density $\rho_\textrm{b}(r)$ and the total gravitational potential $\Phi(r) = \Phi_\textrm{DM}(r) + \Phi_\textrm{b}(r)$ enter. The density $\rho_\textrm{b}(r)$ of the baryonic matter can be inferred from the observed stellar light profile (see Sec.~\ref{Subsec: Baryonic matter content of dwarf galaxies}), while $\rho_\textrm{DM}(r)$ for the dark matter part is model-dependent. In what follows, we will assume the ULDM class of models for $\rho_\textrm{DM}(r)$ and $\Phi_\textrm{DM}(r)$ (see Sec.~\ref{Subsec: Ultralight dark matter}) and calculate the corresponding velocity dispersion (\ref{eq: dispersion general formula}).

\subsection{Observations and fitting procedure}

\begin{figure}[b]
\includegraphics[width=.45\textwidth]{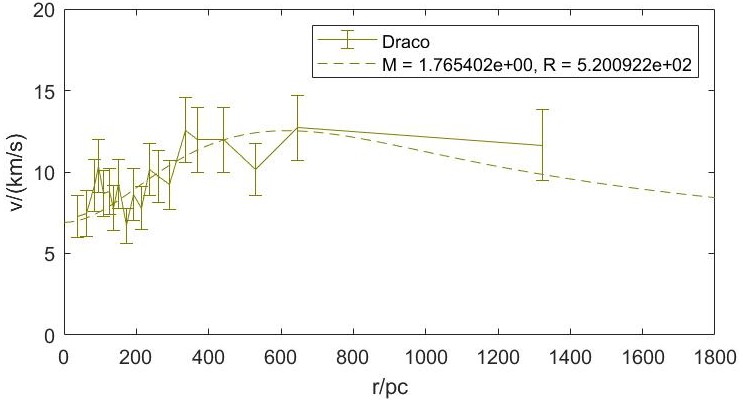}
\includegraphics[width=.45\textwidth]{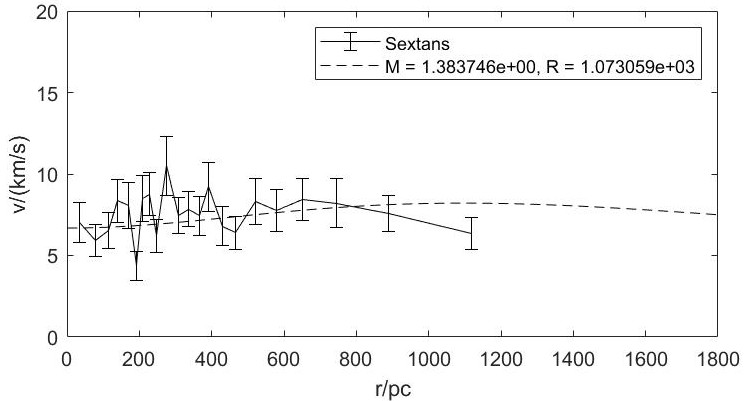}
\caption{\changes{Examples of the velocity dispersion fit for Draco (top) and Sextans (bottom) in the case of attractively interacting ULDM accounting for baryonic matter gravitational potential (see details in Sec.~VB). Here the solid line with the errorbars shows the observed velocity dispersion $v^\textrm{obs}$ from Ref.~\cite{walker2007velocity}, while the dashed line is our fit $v$.} }
\label{Fig: vel disp fit}
\end{figure}

Following Ref.~\cite{walker2007velocity} we consider a sample of seven spheroidal dwarf galaxies orbiting the Milky Way, whose velocity dispersion is known from observations. The observable velocity profile $v^\textrm{obs}(r)$ is compared with the calculated dispersion $v(r) = \sqrt{\bar{v_r^2}}$ from Eq.~(\ref{eq: dispersion general formula}) using the $\chi^2$ fit
\begin{equation}
    \chi^2  = \sum_{j=1}^{\mathcal{N}_g}\sum_{i=1}^{N_j} \frac{[v\changes{_j}(r_i) - v\changes{_j}^\textrm{obs}(r_i)]^2}{\sigma^2_j(r_i)}\, .
    \label{eq: chisquared fit}
\end{equation}
\changes{Here the outer sum ($j = [1,\mathcal{N}_g]$) goes over the $\mathcal{N}_g = 7$ considered galaxies; for each $j$th galaxy the inner sum ($i = [1,N_j]$) goes over the number $N_j$ of the observed values $v_j^\textrm{obs}(r_i)$ of the velocity dispersion.} The $\sigma_j$ is the uncertainty of the observations as given in Ref.~\cite{walker2007velocity}. The theoretical prediction $v\changes{_j}(r_i)$ for the velocity dispersion is defined by Eq.~(\ref{eq: dispersion general formula}) and depends on the unknown ULDM gravitational potential $\Phi_\textrm{DM}(r)$. Equation~(\ref{eq: DM potential}) implies that $\Phi_\textrm{DM}(r)$ is a function of the ULDM mass $M_\textrm{DM}$ and size $R$, where the latter is determined by $M_\textrm{DM}$ and ULDM model parameters $\{m, a_\textrm{s}\}$ (see the mass-radius relations (\ref{eq: first mass-radius relation}) and (\ref{eq: R(M) perturbed by bulge})). Thus, we end up with $v(r)$ which is a function of the halo mass $M_\textrm{DM}$, $m$, and $a_\textrm{s}$.

Then the minimum of \changes{$\chi^2 = \chi^2 (M_\textrm{DM}^1,...,M_\textrm{DM}^{\mathcal{N}_\textrm{g}}, m, a_\textrm{s})$} in the parameter space of the seven halo masses \changes{$M_\textrm{DM}^j$}, the boson mass $m$ and the scattering length $a_\textrm{s}$ corresponds to the best fit of the velocity dispersion observations. The velocity dispersion fits \changes{and the observed data are illustrated by the two examples in Fig.~\ref{Fig: vel disp fit}, other fits} as well as the approach used to find the uncertainty of the fit, can be found in the Supplemental Material \cite{supp}. Our results are given in Sec.~\ref{Sec: Velocity dispersion due to ULDM gravitational potential} where, for simplicity, we neglect the gravitational potential of a small baryonic component. 
The full problem accounting for the gravitational potential of baryonic matter is discussed in Sec.~\ref{Sec: Accounting for baryonic matter gravitational potential}.

\section{Velocity dispersion fit neglecting the baryonic component}
\label{Sec: Velocity dispersion due to ULDM gravitational potential}

Since the considered dwarf spheroidals are DM dominated, we start our discussion by first neglecting the baryonic contribution to the gravitational potential, i.e. $\Phi(r) = \Phi_\textrm{DM}(r)$. 
In this case, a solution of the GPP equations can be approximated by the Gaussian ansatz (\ref{eq: DM density}, \ref{eq: DM potential}) with the unperturbed mass-radius relation $R = R_\textrm{DM}$ (\ref{eq: first mass-radius relation}). Substituting the ULDM gravitational potential (\ref{eq: DM potential}) into Eq.~(\ref{eq: dispersion general formula}) we can find the velocity dispersion (see Appendix~\ref{Appendix: Velocity dispersion for ULDM and baryons}). In what follows, we separately investigate the regimes of repulsively $a_\textrm{s} > 0$ and attractively $a_\textrm{s} < 0$ interacting ULDM.

\subsection{Repulsive interaction}
\label{Sec: Repulsive interaction}

\begin{figure}[htp]
    \centering  \includegraphics[width=.45\textwidth]{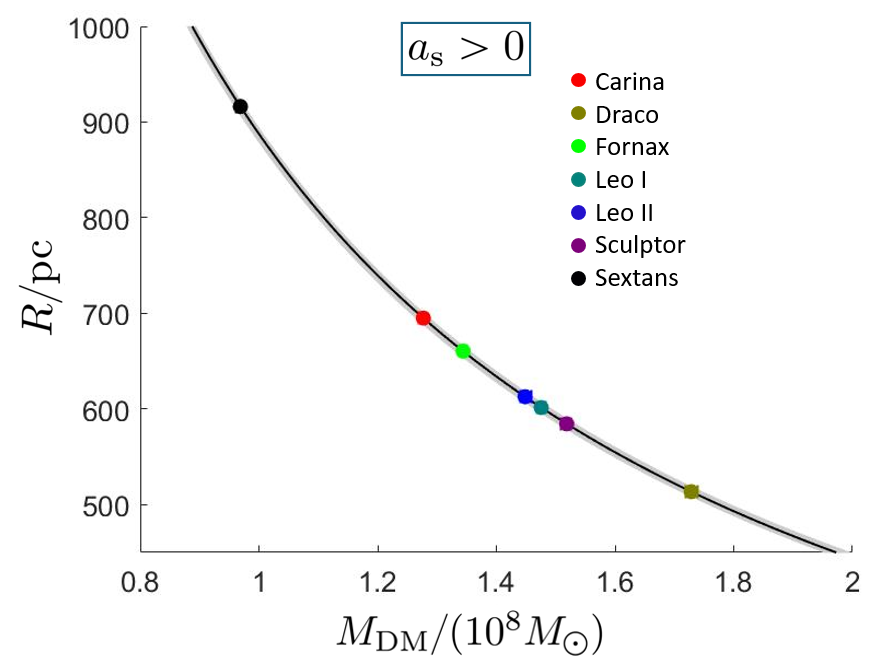}
    \caption{The mass-radius relation for repulsively interacting ULDM. Black solid line shows the best fit, corresponding to the minimum of $\chi^2$-error, while the gray shaded region depicts $M_\textrm{DM}-R$ relations within confidence intervals for $m$ and $a_\textrm{s}$. Colorful points show dSphs with the errorbars corresponding to the confidence intervals for halo masses $M_\textrm{DM}$.}
    \label{Fig3}
\end{figure}

 We start our discussion with the repulsively interacting ULDM $a_\textrm{s} > 0$ in the mass-radius relation (\ref{eq: first mass-radius relation}), illustrated in Fig.~\ref{Fig3}. In this case the best fit to the observations is given by the boson mass $m = 1.902 \times 10^{-22}$~eV (in the range $m \in [1.897, 1.911]\times 10^{-22}$~eV) and the scattering length $a_\textrm{s} = 1.45 \times 10^{-85}$~m (in the range $a_\textrm{s} \in [0, 10^{-79}]$~m).

 \begingroup
\setlength{\tabcolsep}{10pt} 
\renewcommand{\arraystretch}{1.2} 
\begin{table}[h]
\begin{tabular}{ |c|c|c| } 
 \hline 
 Galaxy   & $M_\textrm{DM}/(10^8 M_{\odot})$ & $R$/pc\\  
 \hline
 Carina  &  $1.277^{+0.005}_{-0.009}$ & $695$\\ 
 Draco  & $1.728^{+0.010}_{-0.010}$ & $514$ \\ 
Fornax  & $1.344^{+0.006}_{-0.007}$ & $660$\\
 Leo I & $1.476^{+0.008}_{-0.008}$ & $602$\\
 Leo II  & $1.448^{+0.010}_{-0.008}$ & $613$\\
 Sculptor  & $1.519^{+0.006}_{-0.010}$ & $584$\\
 Sextans   & $0.969^{+0.003}_{-0.008}$ & $916$\\
 \hline
\end{tabular}
\caption{Best fit masses and radii of the ULDM haloes in repulsively interacting ULDM.}
\label{Table rep no bar matter}
\end{table}
\endgroup
 Let us analyze the role of the self-interaction by comparing their contribution to the halo energy $U$ and energy due to the self-gravity $W$ (see Eq.~(\ref{eq: energy compnents})). We find that for the best fit parameters given in Table~\ref{Table rep no bar matter} their ratio is of the order of $U/W \sim 10^{-9}$ for all seven ULDM haloes. Thus, we conclude that the local self-interaction plays a negligibly small role in this case, and ULDM behaves in the same way as the non-interacting fuzzy DM.

\subsection{Attractive interaction}
\label{Subsec: Attractive interaction}

 \begin{figure}[htp]
    \centering  \includegraphics[width=.45\textwidth]{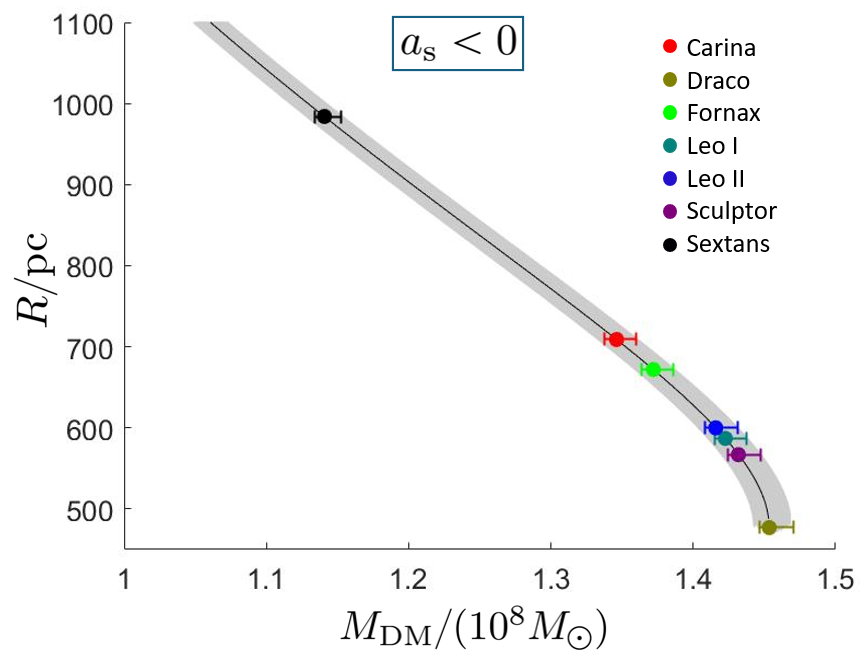}
    \caption{The mass-radius relation for attractively interacting ULDM. The black solid line shows the best fit, corresponding to the minimum of $\chi^2$-error, while the gray shaded region depicts $M_\textrm{DM}-R$ relations within confidence intervals for $m$ and $a_\textrm{s}$. Colorful points show dSphs with the errorbars corresponding to the confidence intervals for halo masses $M_\textrm{DM}$.}
    \label{Fig attractive no bar}
\end{figure}

Now we consider the attractively interacting ULDM $a_\textrm{s} < 0$ in the mass-radius relation (\ref{eq: first mass-radius relation}), illustrated in Fig.~\ref{Fig attractive no bar}. In this case we find that the best fit to the observations is given by the boson mass $m = 1.522 \times 10^{-22}$~eV (in the range $m \in [1.513, 1.532] \times 10^{-22}$~eV) and the scattering length $a_\textrm{s} = -8.66 \times 10^{-78}$~m (in the range $a_\textrm{s} \in [-8.73, -8.52]\times 10^{-76}$~m).

\begingroup
\setlength{\tabcolsep}{10pt} 
\renewcommand{\arraystretch}{1.2} 
\begin{table}[h]
\begin{tabular}{ |c|c|c| } 
 \hline 
 Galaxy   & $M_\textrm{DM}/(10^8 M_{\odot})$ & $R$/pc\\  
 \hline
 Carina  &  $1.346^{+0.014}_{-0.008}$ & $709$\\ 
 Draco  & $1.453^{+0.018}_{-0.007}$ & $477$ \\ 
Fornax  & $1.372^{+0.014}_{-0.008}$ & $672$\\
 Leo I & $1.423^{+0.015}_{-0.008}$ & $587$\\
 Leo II  & $1.416^{+0.015}_{-0.008}$ & $600$\\
 Sculptor  & $1.432^{+0.015}_{-0.008}$ & $567$\\
 Sextans   & $1.141^{+0.012}_{-0.007}$ & $984$\\
 \hline
\end{tabular}
\caption{Best fit masses and radii of ULDM haloes in attractively interacting ULDM.}
\label{Table: attractive no bar}
\end{table}
\endgroup

In contrast to the case of the repulsively interacting ULDM, the interactions play an important role for $a_\textrm{s} < 0$. One can see this directly by comparing the internal energy $U$ with the gravitational energy $W$. Their ratio $U/W = 0.33$ is the biggest for Draco, which is evident from Fig.~\ref{Fig attractive no bar} where the corresponding point $\{M_\textrm{DM}, R\}$ is the closest to the maximum mass $M_\text{max} \approx 1.45 M_\odot$, beyond which the BEC collapses and no stationary state is possible. For the least massive Sextans this ratio is the smallest $U/W = 0.08$, while for other haloes it ranges from $0.16$ to $0.24$.

\section{Accounting for baryonic matter gravitational potential}
\label{Sec: Accounting for baryonic matter gravitational potential}

In this section, we investigate the influence of the baryonic gravitational potential $\Phi_\textrm{b}$ on the velocity dispersion.
Accounting for its contribution in (\ref{eq: dispersion general formula}), we find
\begin{eqnarray}
    \overline{v_r^2} = \frac{1}{
    \rho_\textrm{b}(r)} \int_r^\infty dr' 
    \rho_\textrm{b}(r') \frac{d}{dr'}\left(\Phi_\textrm{DM}(r') + \Phi_\textrm{b}(r')\right) \label{eq: dispersion with baryons}\, .
\end{eqnarray}
The result of the integration is lengthy and can be found in Appendix~\ref{Appendix: Velocity dispersion for ULDM and baryons}. Furthermore, the gravitational potential of baryons affects the shape of the ULDM according to Eqs.~(\ref{eq: R(M) perturbed by bulge}) and (\ref{eq: R(M) perturbed by bulge 2}). 
The mass-radius relation (\ref{eq: R(M) perturbed by bulge}, \ref{eq: R(M) perturbed by bulge 2}) accounts for the shrinking of the ULDM size due to the gravitational attraction from baryonic matter. Note that this new mass-radius relation now depends not only on $m$ and $a_\textrm{s}$ but also on the Plummer radius $b$ and the mass $M_\textrm{b}$ and, therefore, is different for each considered dwarf galaxy with given baryonic matter parameters (see Table~\ref{Table: baryonic parameters}).

\subsection{Repulsive interaction}

\begin{figure}[htp]
    \centering  \includegraphics[width=.45\textwidth]{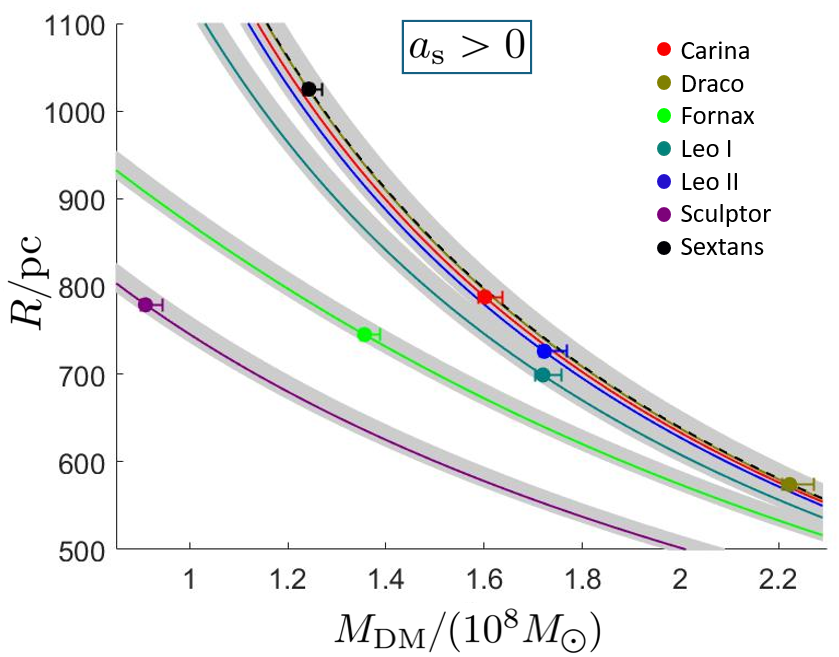}
    \caption{The mass-radius relation for repulsively interacting ULDM. The solid lines show the best fits for different baryonic matter parameters $M_\textrm{b}$ and $b$, while the gray shaded regions correspond to $M_\textrm{DM}-R$ relations within confidence intervals for $m$ and $a_\textrm{s}$. Colorful points depict dSphs with the errorbars corresponding to the confidence intervals for halo masses $M_\textrm{DM}$.}
    \label{Fig5}
\end{figure}

In the $a_\textrm{s} > 0$ case we find that the best fit to the observations is given by the boson mass $m = 1.584 \times 10^{-22}$~eV (in the range $m \in [1.562, 1.596]\times 10^{-22}$~eV) and the scattering length $a_\textrm{s} =  10^{-84}$~m (in the range $a_\textrm{s} \in [0, 4.72 \times 10^{-80}]$~m), as shown in Fig.~\ref{Fig5}.

\begingroup
\setlength{\tabcolsep}{10pt} 
\renewcommand{\arraystretch}{1.2} 
\begin{table}[h]
\begin{tabular}{ |c|c|c| } 
 \hline 
 Galaxy   & $M_\textrm{DM}/(10^8 M_{\odot})$ & $R$/pc\\  
 \hline
 Carina  &  $1.602^{+0.036}_{-0.014}$ & $788$\\ 
 Draco  & $2.223^{+0.050}_{-0.016}$ & $574$ \\ 
Fornax  & $1.356^{+0.031}_{-0.009}$ & $745$\\
 Leo I & $1.720^{+0.037}_{-0.016}$ & $699$\\
 Leo II  & $1.723^{+0.046}_{-0.011}$ & $726$\\
 Sculptor  & $0.910^{+0.036}_{-0.010}$ & $779$\\
 Sextans   & $1.243^{+0.027}_{-0.011}$ & $1025$\\
 \hline
\end{tabular}
\caption{Best fit masses and radii of the ULDM haloes in repulsive ULDM in the presence of baryonic gravitational potential.}
\label{Table: rep with bar}
\end{table}
\endgroup

We note that in this case the interactions are very small. Their relative contribution compared to the gravitational energy is of the order of $U/W \sim 10^{-8}$ for the seven considered galaxies. This agrees with the result from Sec.~\ref{Sec: Repulsive interaction}, where this contribution was also found to be negligibly small. Thus, we conclude that the repulsive self-interaction, if present, is very weak and does not noticeably affect the shape of the ULDM haloes.

\subsection{Attractive interaction}
\label{Subsec: wb Attractive interaction}

\begin{figure}[htp]
    \centering  \includegraphics[width=.45\textwidth]{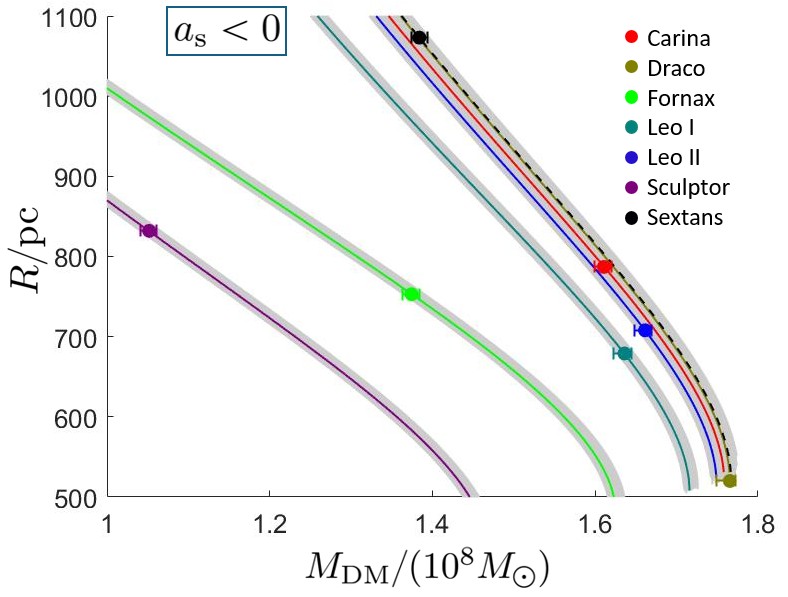}
    \caption{The mass-radius relation for attractively interacting ULDM. The solid lines show the best fits for different baryonic matter parameters $M_\textrm{b}$ and $b$, while the gray shaded regions correspond to $M-R$ relations within confidence intervals for $m$ and $a_\textrm{s}$. Colorful points depict dSphs with the errorbars corresponding to the confidence intervals for halo masses $M$.}
    \label{Fig6}
\end{figure}

Now we consider the attractively interacting ULDM $a_\textrm{s} < 0$ in the mass-radius relation (\ref{eq: R(M) perturbed by bulge}), illustrated by Fig.~\ref{Fig6}. In this case we find that the best fit to the observations is given by the boson mass $m = 1.321 \times 10^{-22}$~eV (in the range $m \in [1.314, 1.327] \times 10^{-22}$~eV) and the scattering length \changes{$a_\textrm{s} =  -6.74 \times 10^{-78}$~m} (in the range $a_\textrm{s} \in [6.72, 6.85]\times 10^{-78}$~m).

\begingroup
\setlength{\tabcolsep}{10pt} 
\renewcommand{\arraystretch}{1.2} 
\begin{table}[h]
\begin{tabular}{ |c|c|c| } 
 \hline 
 Galaxy   & $M_\textrm{DM}/(10^8 M_{\odot})$ & $R$/pc\\  
 \hline
 Carina  &  $1.611^{+0.010}_{-0.012}$ & $787$\\ 
 Draco  & $1.765^{+0.007}_{-0.017}$ & $520$ \\ 
Fornax  & $1.374^{+0.010}_{-0.011}$ & $753$\\
 Leo I & $1.636^{+0.008}_{-0.013}$ & $679$\\
 Leo II  & $1.661^{+0.008}_{-0.013}$ & $708$\\
 Sculptor  & $1.052^{+0.009}_{-0.010}$ & $832$\\
 Sextans   & $1.384^{+0.010}_{-0.010}$ & $1073$\\
 \hline
\end{tabular}
\caption{Best fit masses and radii of the ULDM haloes in attractive ULDM in the presence of baryonic gravitational potential.}
\label{Table: att with bar}
\end{table}
\endgroup

We see that, unlike in the case of repulsively interacting dark matter, the attractive interaction plays an important role in establishing the hydrostatic equilibrium of the BEC. Its contribution is the most prominent for Draco $U/W = 0.33$ and the smallest for Sextans $U/W = 0.08$, while for the other five ULDM haloes $U/W$ it is in the range [0.13,\,0.20] similar to the results obtained in Sec.~\ref{Subsec: Attractive interaction}, where the gravitational potential of baryonic matter was neglected.

\section{ULDM density and discussion}

\subsection{ULDM density}

Let us first compare our results from Secs.~\ref{Sec: Velocity dispersion due to ULDM gravitational potential} and~\ref{Sec: Accounting for baryonic matter gravitational potential} in order to see how strongly the gravitational potential $\Phi_\textrm{b}(r)$ of the baryonic component affects our results. The fits without $\Phi_\textrm{b}(r)$ presented in Tables~\ref{Table rep no bar matter} and~\ref{Table: attractive no bar} and the fits which take into account $\Phi_\textrm{b}(r)$ in Tables~\ref{Table: rep with bar}-\ref{Table: att with bar} do differ. The impact of baryonic matter potential is thus as important as the role of the self-interactions in the ULDM for the fitting of the velocity dispersion. In what follows, we discuss in more detail the results of Sec.~\ref{Sec: Accounting for baryonic matter gravitational potential}, where the baryonic matter potential is accounted for.

\begin{figure}[htp]
    \centering  \includegraphics[width=.45\textwidth]{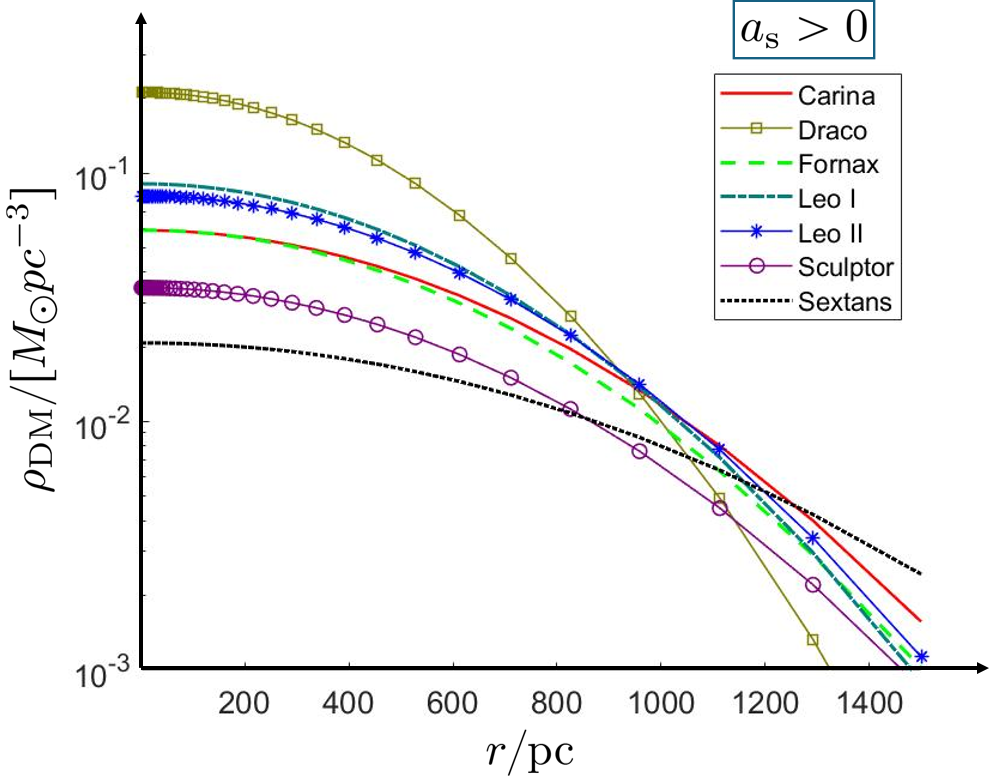}
    \caption{The ULDM density in dSphs as a function of distance for repulsive ULDM whose parameters are given in Table~\ref{Table: rep with bar} (accounting for baryonic component).}
    \label{fig: ULDM density rep}
\end{figure}

\begin{figure}[htp]
    \centering  \includegraphics[width=.45\textwidth]{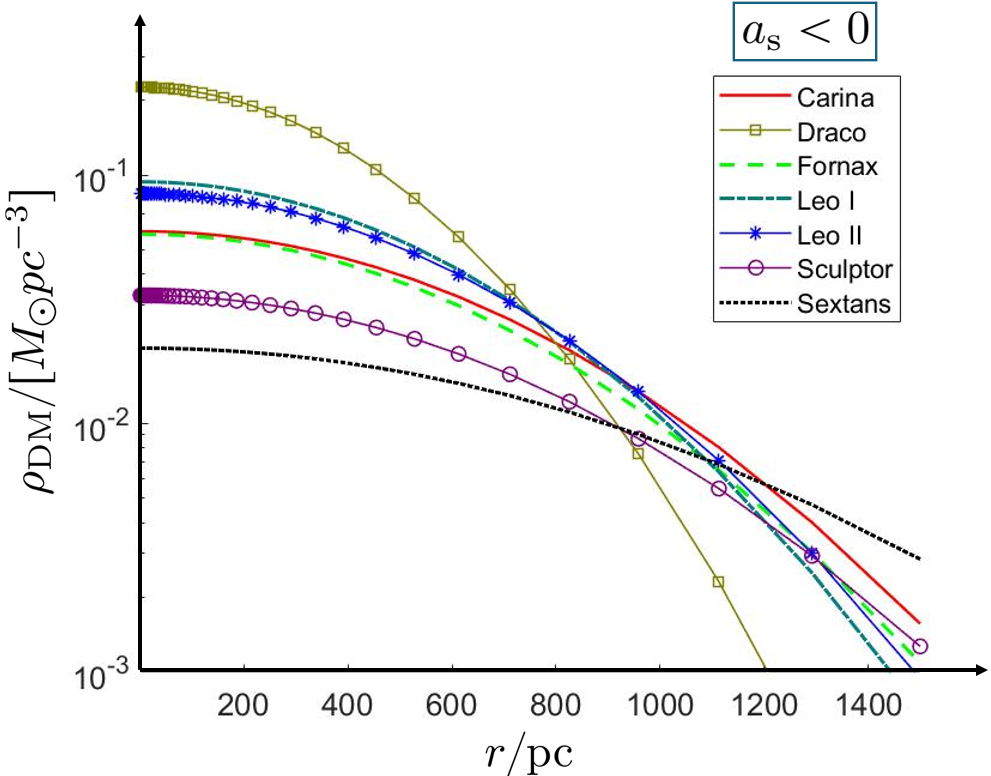}
    \caption{The ULDM density in dSphs as a function of distance for attractive ULDM whose parameters are given in Table~\ref{Table: att with bar} (accounting for baryonic component).}
    \label{fig: ULDM density att}
\end{figure}

We illustrate our findings by plotting the density distributions of ULDM in Figs.~\ref{fig: ULDM density rep} and \ref{fig: ULDM density att}, resulting from the fits of Sec.~\ref{Sec: Accounting for baryonic matter gravitational potential}. We see that while the masses and radii of the ULDM haloes in Tables~\ref{Table: rep with bar} and \ref{Table: att with bar} differ, the density profiles in Figs.~\ref{fig: ULDM density rep} and \ref{fig: ULDM density att} almost resemble each other. This is not surprising since the two fits for repulsively and attractively interacting ULDM reproduce the same observed velocity dispersion data.

\begin{figure}[t]
    \centering  \includegraphics[width=.47\textwidth]{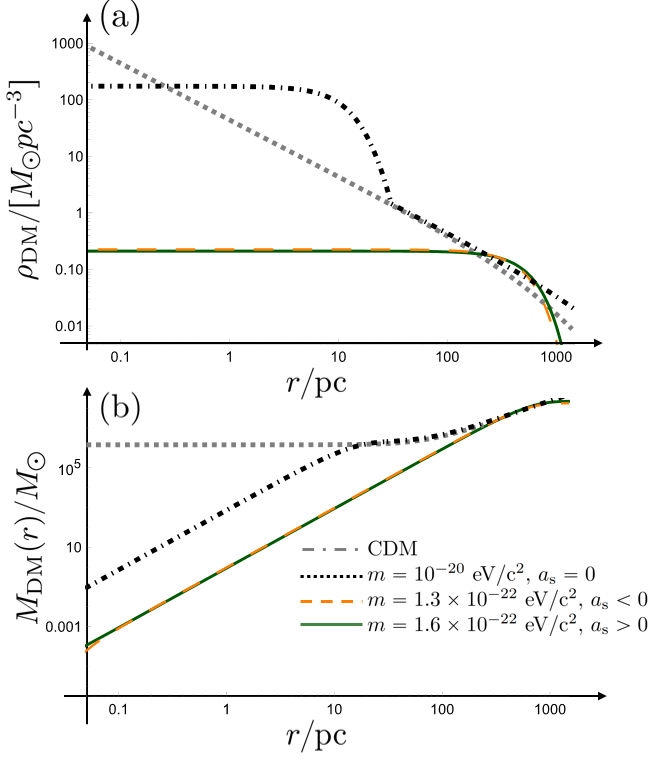}
    \caption{\changes{The DM density $\rho_{\textrm{DM}}(r)$ (a) and the cumulative mass $M_\textrm{DM}(r)/M_\odot$ (b) in Draco as functions of distance $r$ for the four different DM models. The solid green and dashed orange lines show our fits for the repulsively and attractively interacting ULDM with the boson mass $\sim 10^{-22}$~eV/c$^2$. The gray dashed line is the NFW fit from Ref.~\cite{walker2007velocity} assuming standard CDM. The black dash-dotted line shows the combined density of the inner soliton (for $r < 30.7$~pc) and the outer NFW tail (for $r > 30.7$~pc) obtained in Ref.~\cite{goldstein2022viability}}}
    \label{fig: Draco all models}
\end{figure}

Let us discuss the correspondence between our results and the standard CDM model \changes{illustrated for Draco in Fig.~\ref{fig: Draco all models}}. The key difference between CDM and ULDM is that while the former predicts a cuspy density profile in the central region of a DM halo, the latter gives a core with slowly varying density \cite{Ferreira}. For instance, in Ref.~\cite{walker2007velocity}, the same observations were modeled with NFW density profiles assuming CDM. It was concluded that the seven dwarf galaxies are supposed to have masses of the order of $10^{8}M_\odot - 10^{9}M_\odot$, which also holds in our ULDM modeling (see Tables~\ref{Table rep no bar matter}-\ref{Table: att with bar}). The CDM modeling of the velocity dispersion in dwarf galaxies has been also done in Ref.~\cite{hayashi2020diversity}, where it was shown that for the seven dwarfs the DM density is cuspy in the center and tends to a core profile on the outskirts. At distance $r = 150$~pc the DM density is of the order of $0.1$~$M_\odot/$pc$^3$, ranging from $0.05$~$M_\odot/$pc$^3$ for Sextans to $0.26$~$M_\odot/$pc$^3$ for Leo~I. These densities by the order of magnitude agree with the central densities predicted by the ULDM modeling in Figs.~\ref{fig: ULDM density rep} and \ref{fig: ULDM density att}. This can be explained by the fact that both the CDM and ULDM models were fitted to the same observed data. Similarly to Ref.~\cite{hayashi2020diversity} our findings predict that Leo~I and Draco have the highest DM density in the central region, while Sextans has the smallest. \changes{In Fig.~\ref{fig: Draco all models} we also show the ULDM density for the smaller boson mass $m = 10^{-20}$~eV/c$^2$, discussed by \citet{goldstein2022viability} to fit the velocity dispersion observations. In this case, the DM halo mass $10^{9.5}M_\odot$ is distributed over an inner coherent core within $r < 30.7$~pc and the NFW density tail for $r > 30.7$~pc. Since the coherent core constitutes only $\sim 0.1 \%$ mass of the halo, we see that the ULDM is similar to the standard CDM except for the presence of a small core in the center.}

\subsection{Discussion}

In this subsection, we discuss the physical implications and limitations of the fitted ULDM models, comparing our results with existing astrophysical constraints and evaluating their consistency with previous studies. We note that the model of self-interacting ULDM with the boson mass $\sim 10^{-22}$~eV may face some challenges other than the considered fitting of velocity dispersion observations.

For instance, ULDM has its own velocity dispersion $v_\textrm{DM}$ and the corresponding de Broglie wavelength $\lambda_\textrm{dB} = \hbar/(mv_\textrm{DM})$. The expression for $v_\textrm{DM}$ can be found in Appendix~\ref{Appendix: Velocity dispersion for ULDM and baryons}. In our further considerations, we take $v_\textrm{DM} \sim 10$~km/s (in the range between $12$ and $20$ km/s for different ULDM haloes and different models) as an estimate of the velocity dispersion in the central region of the halo. For this value and $m \sim 10^{-22}$~eV, we find $\lambda_\textrm{dB} \sim 10^3$~kpc which is $10^3-10^4$ times bigger than the ULDM halo size $R$. For ultra-faint dwarf galaxies, which are smaller than the considered seven dSphs, with a typical size $50$~pc and velocity dispersion $v \sim 5$~km/s, one obtains the ratio $\lambda_\textrm{dB}/R = 10^5-10^6$ \cite{teodori2025ultralightdarkmattersimulations}.

The aforementioned ratio $\lambda_\textrm{dB}/R$ plays an important role in the dynamics of an ULDM halo, since it sets the relative length scale where the density fluctuations of BEC occur \cite{may2025updated}. These fluctuations in a BEC core can be triggered by a range of astrophysical processes, altering the ULDM density distribution on gigayear timescales \cite{salasnich2025collective, PhysRevD.103.023508}. These density fluctuations produce gravitational perturbations which affect the dynamics of stars in the galaxy and lead to dynamical heating \cite{teodori2025ultralightdarkmattersimulations, dalal2022excluding, may2025updated,salasnich2025collective, PhysRevD.103.023508}. According to the findings of Refs.~\cite{teodori2025ultralightdarkmattersimulations, dalal2022excluding, may2025updated}, the dynamical heating of stars becomes critical for $\lambda_\textrm{dB}/R \gtrsim 10^2$ when density fluctuations exhibit sufficiently long-range coherence compared to the halo size. The requirement $\lambda_\textrm{dB}/R \gtrsim 10^2$ sets the limit on ULDM mass, which was found in Ref.~\cite{teodori2025ultralightdarkmattersimulations} to be $m > 5 \times 10^{-21}$~eV. Even more severe constraints $m > 3 \times 10^{-19}$~eV and $m > 8 \times 10^{-18}$~eV can be obtained from the analysis of the ultra-faint dwarf galaxies \cite{dalal2022excluding,may2025updated}. Therefore, this argument could exclude the considered model of ULDM boson with mass $m \sim 10^{-22}$~eV based on the observations of both dwarf and ultra-faint dwarf galaxies. However, we note that the recent spectroscopic analysis of \textit{Ursa Major III / UNIONS 1} by \citet{cerny2025no} suggests that this system may be a dark-matter–free star cluster rather than a dark-matter–dominated galaxy as assumed by \citet{may2025updated}. Consequently, the stringent ULDM particle mass limits ($m > 8 \times 10^{-18}~\mathrm{eV}$) derived from this object are now debated, as their validity depends on the still-uncertain nature of \textit{Ursa Major III / UNIONS 1}.

The constraints obtained in Refs.~\cite{teodori2025ultralightdarkmattersimulations, dalal2022excluding, may2025updated} are based on the model of fuzzy DM, i.e., ULDM in the absence of short-range interactions. In the case of repulsively interacting ULDM our findings show that the interactions are negligibly small and the same limits for the boson mass should apply. Attractive self-interactions (see Sec.~\ref{Subsec: wb Attractive interaction}), which produce effects comparable to the self-gravity, could influence these constraints \cite{may2025updated}. We note that for the two fitted ULDM models $\{m, a_s\}$ in the two cases of $a_s > 0$ and $a_s < 0$ (see Tables~\ref{Table: rep with bar} and~\ref{Table: att with bar}, respectively) the velocity dispersion $v_\textrm{DM}$ does not differ significantly and, therefore, the short-range self-interaction cannot resolve this tension. We also note that the model of $a_\textrm{s} < 0$ may contradict with the Lyman~$\alpha$ forest observations \cite{irvsivc2024unveiling}. The short-range interactions, considered in Sec.~\ref{Subsec: wb Attractive interaction}, are characterized by the ratio $\lambda = 16\pi/3 \times a_\textrm{s} / \lambda_\textrm{c} \sim 10^{-91}$ of the s-wave scattering length $a_\textrm{s}$ and the Compton wavelength $\lambda_\textrm{c} = \hbar/(mc)$, which is too big to explain the small-scale power spectrum of ultra-faint galaxies \cite{may2025updated}.

We see that while the velocity dispersion observations favor ULDM with mass $\sim 10^{-22}$~eV, some other observations exclude this possibility. In order to strengthen this conclusion and fully exclude the ULDM model, one would need to provide a detailed modeling of baryonic component, incoherent structures in ULDM \cite{Liu_2023, indjin2025fuzzy}, black holes, and other relevant effects for a realistic galaxy, which poses a challenging task \cite{sales2022baryonic}. For instance, if an ultra-faint dwarf would host a black hole (BH), the latter would produce additional gravitational attraction, thus, decreasing $\lambda_\textrm{dB}$. This possibility cannot be excluded for the observed ultra-faint galaxies so far, since not all such BHs would produce the requisite accretion signatures to
be detected \cite{reines2022hunting}. For example, the observed stellar dynamics in the ultra-faint dwarf \textit{Segue~I} favors the presence of a supermassive BH \cite{lujan2025modeling}, which was not accounted by \citet{dalal2022excluding} to constrain ULDM. This was not considered in the present study and in the constraints obtained in Refs.~\cite{teodori2025ultralightdarkmattersimulations, dalal2022excluding, may2025updated}, while it can have an important impact on the modeling of ULDM halos \cite{PhysRevD.111.023006}. 

Moreover, in Refs.~\cite{dalal2022excluding,teodori2025ultralightdarkmattersimulations, may2025updated},
it was assumed that ULDM is described by a fully coherent wave function which satisfies the GPP equations (\ref{GP equation}) and (\ref{Poisson equation}). Hence, the limitations for boson mass, that come from the appearance of ULDM density fluctuations on the $\lambda_\textrm{dB}$ scale, rely on the assumption that this coherence is dynamically preserved in the whole simulated volume. This is not necessarily the case, having that the correlations in the self-gravitating BEC \cite{Liu_2023, indjin2025fuzzy} drop with distance and on the outskirts the ULDM is in a turbulent, rather than a coherent state. In addition, unlike CDM, ULDM can form stable vortex structures with quantized superfluid flow \cite{nikolaieva2021stable}, which may noticeably influence the DM density distribution and dynamics of stars in a dwarf galaxy \cite{PhysRevD.111.023006, PhysRevD.108.023503}.

\section{Conclusions}

Investigating velocity dispersion curves of seven dwarf spheroidals in the context of the ULDM class of models with local self-interaction, we found that the velocity dispersion in these galaxies is determined mainly by the dominating ULDM gravitational potential and, therefore, allows us to efficiently constrain the mass and shape of DM haloes. For the considered seven galaxies, this fixes the mass-radius relation, which uniquely determines the corresponding boson mass $m$ and its s-wave scattering length $a_\textrm{s}$. This approach made it possible to deduce the most optimal ULDM halo masses, radii, and parameters $\{m, a_\textrm{s}\}$ by fitting the data from observations. We also find that the gravitational attraction of the baryonic component in each dwarf galaxy makes the solitonic ULDM halo more compact and modifies the ULDM mass-radius relation. Accounting for this effect as well as the contribution of the baryonic potential to the velocity dispersion, we consider the role of the baryonic component in Sec.~\ref{Sec: Velocity dispersion due to ULDM gravitational potential} and Sec.~\ref{Sec: Accounting for baryonic matter gravitational potential}.

We conclude that the velocity dispersion observations in the dwarf galaxies are compatible with the ULDM models considered here and, in particular, the dynamics imposed by a soliton. The main results of our work are parameters of ULDM models and masses of ULDM haloes presented in Tables~II-IV. We see that the masses, radii, and relations between them for the dwarf spheroidals depend on whether ULDM interacts repulsively or attractively. Our results suggest that only a very weak repulsive interaction $a_\textrm{s} \lesssim 10^{-80}$~m is in agreement with the velocity dispersion observations allowing for non-interacting ULDM with $m \approx 1.6 \times 10^{-22}$~eV. In contrast, if the self-interaction in ULDM is attractive, $a_\textrm{s} \sim -10^{-78}$~m, then we have a smaller boson mass $m \approx 1.3 \times 10^{-22}$~eV and a completely different mass-radius relation. These results imply that the presence of self-interactions in ULDM plays a significant role in the modeling of observations. Our results for densities of the seven dSphs show qualitative agreement with Refs.~\cite{walker2007velocity, hayashi2020diversity}, where the velocity dispersion of the dwarf galaxies was fitted by CDM. The CDM density profiles have a cusp in the center and at larger distances $\sim 150$~pc tend to a core-like profile with densities similar to those we find.

Our findings predict that the ULDM haloes of dwarf galaxies have ULDM velocity dispersion $\sim 10$~km/s, which implies that the corresponding de Broglie wavelength is $10^3-10^4$ times larger than the size of the halo. In this case, the density fluctuations of BEC lead to the dynamical heating of the baryonic component that contradicts observations \cite{teodori2025ultralightdarkmattersimulations, dalal2022excluding, may2025updated}. Moreover, the stars kinematics in Leo~II implies the constraint $m > 2.2 \times 10^{-21}$~eV on the boson mass \cite{zimmermann2025dwarf}. This is indirectly supported by our results because we found the biggest disagreement between the ULDM fit and the observable velocity dispersion in Leo~II. This hints that Leo~II may be an optimal candidate to exclude ULDM from the stellar kinematics perspective.

Our results clarify the role of the self-interaction and baryonic matter in fitting of the velocity dispersion observations of dwarf galaxies. This contributes to the previous research \cite{goldstein2022viability, Ferreira, calabrese2016ultra, teodori2025ultralightdarkmattersimulations, dalal2022excluding, may2025updated,zimmermann2025dwarf}, which focused on the observationally based constraints on ULDM models. While these findings so far exclude ULDM, still, they do not account for all the possible effects, which may play role in a realistic scenario. The realistic modeling of a galaxy is a computationally challenging task and would require a detailed dynamical modeling of baryonic matter, collective excitations, black holes, tidal stripping, superfluid flows and NFW tails. This poses an open question for our analysis and should be further addressed to provide more insight in the small-scale nature of DM.

\vspace{7mm}

\centerline{\bf Acknowledgements}
\vspace{2mm}
The authors acknowledge Srikanth Nagesh, Andrii Momot and Bohdan Hnatyk for fruitful discussions. 
The work of E.V.G. and Y.R. was support by the Swiss National Science Foundation through the Ukrainian-Swiss Joint research project "Cosmic waltz of baryonic and ultralight dark matter: interaction and dynamical interplay" (grant No. IZURZ2{\_}224972). K.K. acknowledges funding by the Deutsche Forschungsgemeinschaft (DFG, German Research Foundation) under Germany’s Excellence Strategy—EXC 2123 QuantumFrontiers—390837967.

\appendix

\section{Mass-radius relation with gravitational potential of ULDM and baryons}
\label{Appendix: Mass-radius relation for ULDM with baryons}

In Ref.~\cite{chavanis2011mass} it was found that the mass-radius relation $R(M)$ for an ULDM soliton is given by Eq.~(\ref{eq: first mass-radius relation}). In this Appendix we discuss how this mass-radius relation is modified in the case when ULDM is subject to the additional gravitational potential of the baryonic matter~(\ref{eq: phib}).

Following Ref.~\cite{chavanis2011mass}, we find the energy of ULDM 
\begin{eqnarray}
    E &=& \frac{3}{4} \frac{\hbar^2 M_\textrm{DM}}{m^2 R^2} + \frac{a_\textrm{s} \hbar^2 M_\textrm{DM}^2}{\sqrt{2\pi} m^3 R^3} - \frac{GM_\textrm{DM}^2}{\sqrt{2\pi} R} \nonumber \\
    &-& \frac{4G M_\textrm{DM} M_\textrm{b}}{\sqrt{\pi} R} \int_{0}^{+\infty}dx \frac{x^2 e^{-x^2}}{\sqrt{x^2 + (b/R)^2}}\, ,
\end{eqnarray}
where the last term is the energy $\int \rho_\textrm{DM}(r)\Phi_\textrm{b}(r) d\mathbf{r}$ of the gravitational interaction between ULDM and baryonic matter. The mass radius relation is then determined by 
\begin{eqnarray}
    \frac{\partial E}{\partial R} &=& -\frac{3}{2} \frac{\hbar^2 M_\textrm{DM}}{m^2 R^3} - \frac{3 a_\textrm{s} \hbar^2 M_\textrm{DM}^2}{\sqrt{2 \pi} m^3 R^4} + \frac{GM_\textrm{DM}^2}{\sqrt{2\pi}R^2} \nonumber \\
    &+& \frac{3G  M_\textrm{DM} M_\textrm{b}}{2 R^2}\textrm{U}\left[\frac{3}{2}, 0, \left(\frac{b}{R}\right)^2\right]= 0\, \label{eq: energy minimum},
\end{eqnarray}
where U[$x$, $y$, $z$] stands for the Tricomi confluent hypergeometric function (\ref{eq: confluent U}). The latter equation can be solved in the general case only numerically, however, in the regime $M_\textrm{b} \ll M_\textrm{DM}$, one can obtain an approximate analytical solution. Indeed, in this case, the perturbed size of ULDM only slightly differs from the unperturbed result (\ref{eq: first mass-radius relation}), which we denote as $R_\textrm{DM}$. This allows us to approximate $R$ by $R_\textrm{DM}$ in the argument of $U$ function in the last term of Eq.~(\ref{eq: energy minimum}) and obtain the solution
\begin{eqnarray*}
    R &=& \frac{3\sqrt{2\pi}}{4} \tilde{R}_Q \left(1 \pm \sqrt{1 + \frac{8}{3 \pi} \frac{M_\textrm{DM}}{m} \frac{a_\textrm{s}}{\tilde{R}_Q} }\right)\\
    \tilde{R}_Q &=& \frac{R_Q}{\left(1 + M_\textrm{b}/M_\textrm{DM} \times f (b/R_\textrm{DM})\right)} \, ,
\end{eqnarray*}
where the radius $R$ is now defined by a new length scale $\tilde{R}_Q$, which in the absence of baryonic matter equals $R_Q$ as given by Eq.~(\ref{eq: first mass-radius relation}).
We see that the latter result shows the decrease of the ULDM radius due to gravitational attraction from the baryonic matter. This approximation quantitatively agrees within one percent difference compared to the numerical solution to Eq.~(\ref{eq: energy minimum}) and is used in our calculations.

\section{Velocity dispersion for ULDM and baryons}
\label{Appendix: Velocity dispersion for ULDM and baryons}

We can write down the expression for velocity dispersion (\ref{eq: dispersion with baryons}) as a sum of two integrals
\begin{equation}
    \bar{v_r^2}(r) = \frac{I_\textrm{DM}(r) + I_\textrm{b}(r)}{\rho_\textrm{b}(r)}\label{eq: baryon vel disp}\, ,
\end{equation}
where the integrals $I_\textrm{DM}$ and $I_\textrm{b}$ above are the functions of the BEC size $R$, distance $y = r/R$, and Plummer radius $\beta = b/R$ in the units of $R$. The function 
\begin{equation*}
    I_\textrm{b} = \int_r^\infty dr' \rho_\textrm{b}(r')\frac{d \Phi_\textrm{b}}{dr'} = \frac{G M_\textrm{b}^2 \beta^2}{8\pi R^4}\frac{1}{u^6}\, 
\end{equation*}
describes the contribution of baryons and
\begin{eqnarray*}
    I_\textrm{DM} &=& \int_r^\infty dr' \rho_\textrm{b}(r')\frac{d \Phi_\textrm{DM}}{dr'} = \frac{G M_\textrm{DM} M_\textrm{b}}{4 \pi^{3/2} \beta^4 R^4} \frac{e^{-y^2}}{y u^3} \\
    &\times&\left[-4 \sqrt{\pi} \left(\beta ^4-2 \beta ^2+2\right) e^{u^2} u^3 y \text{erf}(u) \right.
\\ &+&\sqrt{\pi} e^{y^2} \text{erf}(y) \left(3 \beta ^4+8 y^4+12 \beta ^2 y^2\right)\\
&+& 2 y \left(-2 \beta ^6+5 \beta ^4-4 \sqrt{\pi } u^3 e^{y^2}+2 \sqrt{\pi } \left(\beta ^2-2\right) \beta ^2 e^{u^2} u^3 \right. \\&+& \left. \left. 4 \sqrt{\pi }  e^{u^2} u^3-2 \left(\beta ^2-2\right) \beta ^2 y^2\right)\right]
\end{eqnarray*}
is the contribution due to the ULDM gravitational potential. Here we also introduced notation $u = \sqrt{y^2 + \beta^2}$ for brevity. The expressions above determine our theoretical prediction for the velocity dispersion of baryons.

While the velocity dispersion of DM $v_\textrm{DM} = \left(\overline{v_\textrm{DM}^2}\right)^{1/2}$ is not directly observable, it is relevant in the analysis of such properties of a halo as its effective termperature and de Broglie length \cite{chavanis2019predictive, may2025updated}. It can be evaluated as
\begin{eqnarray}
    \overline{v_\textrm{DM}^2} &=&  \frac{1}{\rho_\textrm{DM}(r)}\int_r^\infty dr' \rho_\textrm{DM}(r')\frac{d \Phi_\textrm{DM}}{dr'} = GM_\textrm{DM} \label{eq: Gaussian velocity dispersion} \\
    &\times& \left(\frac{\textrm{erf}[r/R]}{r} + \frac{\sqrt{\pi}}{2R}e^{r^2/R^2} \left(-1 + (\textrm{erf}[r/R])^2\right)\right).
    \nonumber
\end{eqnarray}

\bibliography{main}

\end{document}